\documentclass[english]{cccconf} 

\usepackage[comma,numbers,square,sort&compress]{natbib}
\usepackage{epstopdf}
\usepackage{fancyhdr}
\usepackage{indentfirst}
\usepackage{amsmath}
\usepackage{amssymb}
\usepackage{graphicx}
\usepackage{color}
\usepackage{latexsym}
\usepackage{amsfonts}
\usepackage{amssymb,amsmath,amsfonts}
\usepackage{amsmath,mathrsfs,bm,url,times}
\usepackage{caption}
\usepackage{subcaption}
\usepackage{textcomp}
\newtheorem{assumption}{Assumption}

\newtheorem{theorem}{Theorem}
\newtheorem{lemma}{Lemma}

\newtheorem{remark}{Remark}

\newcommand{\R}{{\mathbb R}}
\DeclareMathOperator{\sat}{sat}
\DeclareMathOperator{\argmax}{argmax}
\DeclareMathOperator{\diag}{diag}
\DeclareMathOperator{\SCC}{SCC}
\DeclareMathOperator{\Deg}{Deg}
\DeclareMathOperator{\rank}{rank}
\allowdisplaybreaks

\begin{document}
\title{\LARGE \bf
Event-Triggered Control for Consensus of Multi-Agent Systems with Nonlinear Output and Directed Topologies
}


\author{Xinlei Yi\aref{kth}, Shengjun Zhang\aref{unt}, Tao Yang\aref{unt}, Junfeng Wu\aref{jfw}, Karl H. Johansson\aref{kth}
}

\affiliation[kth]{School of Electrical Engineering and Computer Science, KTH Royal Institute of Technology, 100 44, Stockholm, Sweden
        \email{\{xinleiy, kallej\}@kth.se}}
\affiliation[unt]{The Department of Electrical Engineering, University of North Texas, Denton, TX 76203 USA
        \email{ShengjunZhang@my.unt.edu, Tao.Yang@unt.edu}}

\affiliation[jfw]{The College of Control Science and Engineering, Zhejiang University, 310027, Hangzhou, P. R. China
        \email{jfwu@zju.edu.cn}}

\maketitle

\begin{abstract}\label{s:Abstract}
We propose a distributed event-triggered control law to solve the consensus problem for multi-agent systems with nonlinear output. Under the condition that the underlying digraph is strongly connected, we propose some sufficient conditions related to the nonlinear output function and initial states to guarantee that the event-triggered controller realizes consensus. Then the results are extended to the case where the underlying directed graph contains a directed spanning tree.
These theoretical results are illustrated by numerical simulations.
\end{abstract}


\footnotetext{This work is supported by the
Knut and Alice Wallenberg Foundation, the  Swedish Foundation for Strategic Research, the Swedish Research Council, and the Ralph E. Powe Junior Faculty
Enhancement Award for the Oak Ridge Associated Universities (ORAU).}

\section{INTRODUCTION}
\label{sec:intro}
Consensus in multi-agent systems has been extensively studied partially due to its wide applications.
Various distributed consensus protocol have been developed, see, e.g., \cite{olfati2004consensus,ren2007information,Liu2011consensus}.
In these protocols, each agent updates its state based on its own state and the states of its
neighbors in such a way that the final states of all agents converge to a common value.
However, physical systems are subject to constraints such as input and output saturation, digital communication channels, and sensors constraints.
These constraints lead to nonlinearity in the closed-loop dynamics.
Thus the behavior of each agent is affected and special attention to these constraints needs to be taken in order to understand their influence on the convergence properties.
Here we list some representative examples of such constraints.
For example, \cite{yang2014global} studies global consensus for discrete-time multi-agent systems with input saturation constraint; \cite{meng2013global} considers the leader-following consensus problem for multi-agent systems subject to input saturation; \cite{lim2016consensus} and \cite{wang2016conditions} investigate necessary and sufficient initial conditions for achieving consensus in the presence of output saturation. Multi-agent systems with general nonlinear dynamics have also been extensively studied, see, e.g., \cite{liu2009consensus,wei2018nonlinear}.

It may be impractical to require continuous communication in physical
applications, as agents are equipped with embedded microprocessors with limited resources to transmit and collect information.
Event-triggered control was introduced partially to tackle this problem \cite{aastrom1999comparison,tabuada2007event}.
The control in event-triggered control is often piecewise constant between
the triggering times.
The triggering times are determined implicitly by the event conditions.
Many researchers studied event-triggered control for multi-agent systems recently, e.g., \cite{dimarogonas2012distributed,seyboth2013event,meng2013event,yi2016distributed,yi2017pull,yi2018dynamic}.
Key challenges are how to design the control law, to determine the event threshold, and to avoid Zeno behavior.
For continuous-time multi-agent systems, Zeno behavior is that there are infinite number of triggers in a
finite time interval \cite{johansson1999regularization}.
One more challenge is to realize the event-triggered controller in a distributed way since on the one hand for the centralized controller, it normally requires a large amount of agents take action in a synchronous manner and it is not efficient; on the other hand for some decentralized controller, it requires a priori knowledge of some global network parameters, for example, in \cite{seyboth2013event} the smallest positive eigenvalue of the Laplacian matrix needs to be known in advance.

In this paper, we consider event-triggered control for consensus of multi-agent systems with nonlinear output and directed topologies.
We first consider the case that the underlying digraph is strongly connected.
In this case, we provide sufficient conditions related the nonlinear function and the initial states to achieve consensus.
Then, we study the case that the underlying digraph has a directed spanning tree and show that consensus is still achieved if more conditions are satisfied.
The contributions of this paper are as follows: we consider nonlinear output; the underlying graph considered in this paper is directed and has a spanning tree; our event-triggered control law is distributed in the sense that it does not require any a priori knowledge of global network parameters;  and we show that it is free from Zeno behavior by proving that the triggering time sequence of each agent is divergent.
In other words, we extend our previous results \cite{yi2017event} which considered multi-agent systems with output saturation under undirected graphs to general nonlinear output under directed graphs.


The rest of this paper is organized as follows. \mbox{Section \ref{secpreliminaries}} introduces the preliminaries. The main results are stated in Section \ref{secmain}. Simulations are given in Section \ref{secsimulation}. Finally, the paper is concluded in Section \ref{secconclusion}.

\noindent {\bf Notions}: $\|\cdot\|$ represents the Euclidean norm for
vectors or the induced 2-norm for matrices. ${\bf 1}_n$ denotes the column one
vector with dimension $n$. $I_n$ is the $n$ dimensional identity matrix. $[n]$ represents the set $\{1,\dots,n\}$ for any positive integer $n$. $\rho(\cdot)$ stands for
the spectral radius for matrices and $\rho_2(\cdot)$ indicates the minimum
positive eigenvalue for matrices having positive eigenvalues. Given two
symmetric matrices $M$ and $N$, $M>N$ ($M\ge N$) means that $M-N$ is positive
definite (positive semi-definite). For a matrix $A$, $A^\top$ denotes its transpose and $\rank(A)$ is its rank. Given a vector $[x_1,\dots,x_n]^\top\in\mathbb{R}^n$, $\diag([x_1,\dots,x_n])$ is a diagonal matrix with the $i$-th diagonal element being $x_i$. The notation $A\otimes B$ denotes the Kronecker product
of matrices $A$ and $B$. Given a vector $s=[s_1,\dots,s_n]^\top\in\mathbb{R}^n$, define the component operator $c_l(s)=s_l,l\in[n]$.

\section{PRELIMINARIES}\label{secpreliminaries}
In this section, we present some definitions in algebraic graph theory \cite{mesbahi2010graph} and the considered multi-agent system.

\subsection{Algebraic Graph Theory}

Let $\mathcal G=(\mathcal V,\mathcal E, A)$ denote a weighted directed graph (digraph) with the set of agents (vertices) $\mathcal V =\{v_1,\dots,v_n\}$, the set of links (edges) $\mathcal E\subseteq \mathcal V \times \mathcal V$, and the weighted adjacency matrix
$A =(a_{ij})$ with nonnegative elements $a_{ij}$. A link
of $\mathcal G$ is denoted by $(v_i,v_j)\in \mathcal E$ if there is
a directed link from agent $v_i$ to agent $v_j$ with weight $a_{ji}>0$, i.e., agent $v_i$ can send
information to agent $v_j$. The adjacency elements associated with the links of the graph
are positive, i.e., $(v_i,v_j)\in \mathcal E\iff a_{ji}>0$. It is assumed that $a_{ii}=0$ for all $i\in[n]$.
The in-degree of agent $v_i$ is defined as $
\deg^{\text{in}}_i=\sum\limits_{j=1}^{n}a_{ij}$.
The in-degree matrix of $\mathcal G$ is defined as
$\Deg^{\text{in}}=\diag([\deg^{\text{in}}_1, \dots, \deg^{\text{in}}_n])$.
The (weighted) Laplacian
matrix associated with $\mathcal G$ is defined as $L=\Deg^{\text{in}}-A$.
A directed path from agent $v_0$ to agent $v_k$ is a directed graph
with distinct agents $v_0,...,v_k$ and links $(v_i,v_{i_1}),(v_{i_1},v_{i_2}),\dots,(v_{i_{k-1}},v_{i_k}),(v_{i_k},v_j)$.
A digraph $\mathcal G$ is strongly connected if for any two
distinct agents $v_i$ and $v_j$, there exists a directed path from $v_i$ to
$v_j$.
A digraph $\mathcal G$ has a directed spanning tree if there exists
one agent such that there exists a directed path from this agent to any other agent.

By Perron-Frobenius theorem \cite{lu2006new} (for more detail and proof, see
\cite{lu2007new}), we have
\begin{lemma}\label{lemmaL}
If $L$ is the Laplacian matrix associated with a digraph $\mathcal G$ that has a directed spanning tree,
then $\rank(L)=n-1$, and there is a nonnegative vector
$\xi^{\top}=[\xi_{1},\dots,\xi_{n}]$ such that $\xi^{\top} L=0$ and
$\sum_{i=1}^{n}\xi_{i}=1$. Moreover, if $\mathcal G$ is strongly connected, then $\xi_i>0,i\in[n]$ and $\Xi L+L^{\top}\Xi$ is a symmetric matrix with all row sums equal to zeros and has zero eigenvalue with algebraic dimension one, where $\Xi=\diag(\xi)$.
\end{lemma}


\subsection{Multi-Agent Systems with Nonlinear Output}
We consider a set of $n$ agents that are modelled as a single integrators with output saturation:
\begin{align}
\begin{cases}
\dot{x}_i(t)=u_i(t),\\
y_i(t)=g(x_i(t))
\end{cases}
,~i\in[n], t\ge0,\label{system}
\end{align}
where $x_i(t),u_i(t),y_i(t)\in\mathbb{R}^p$ are the state, the control input, and the measured output of agent $v_i$, respectively, and $g: \mathbb{R}^p \rightarrow \mathbb{R}^p$ is a nonlinear function defined as
\begin{align*}
g(s)=[g_{1}(s_1),\dots,g_{p}(s_p)]^\top
\end{align*}
with $s=[s_1,\dots,s_p]^\top\in\mathbb{R}^p$.

We now state some mild assumptions on the nonlinear function $g$ as follows.

\begin{assumption}\label{assnonlinearg}
For all  $l\in [p]$,
\begin{itemize}
  \item[1)] $g_{l}(x)$ is continuous for all $x\in\mathbb{R}$;
  \item[2)] $g_{l}(x)$ is nondecreasing on $\mathbb{R}$ and there exists $h>0$ such that  $g_{l}(x)$ is strictly increasing on $[-h,h]$;
  \item[3)] $g_{l}(x)=0$ if and only if $x=0$;
  \item[4)] $g_{l}(x)$ is locally Lipschitz, i.e., for any compact set $\mathcal{S}\subseteq\mathbb{R}$,  there exist a constant $K(\mathcal{S})>0$, such that $|g_l(x)-g_l(y)|\le K(\mathcal{S})|x-y|,~\forall x,y\in\mathcal{S}$;
  \item[5)] there exists $\varrho>0$ such that  $|g_{l}(x)-g_{l}(y)|\ge\varrho|x-y|$ for all $x,y\in[-h,h]$.
\end{itemize}
\end{assumption}
Note that Assumption~\ref{assnonlinearg} is satisfied for some commonly used functions, for example the identity mapping and saturation function
\begin{align*}
\sat_h(x)=
\begin{cases}
h,~&\text{if}~x\ge h\\
x,~&\text{if}~|x|<h\\
-h,~&\text{if}~x\le -h
\end{cases}.
\end{align*}

\begin{remark}
For the ease of the presentation, we study the case where all the agents have the same $h$ and $\varrho$ in Assumption~\ref{assnonlinearg}. However, the analysis in this paper is also valid for the case where the agents have different $h$ and $\varrho$.
\end{remark}

The following lemma is useful. The proof is straightforward. We thus omit the details here.
\begin{lemma}\label{lemma3}
Suppose the first two parts in Assumption~\ref{assnonlinearg} hold.
For any constant $a$, define $G_l(x)=\int_{a}^{x}(g_l(s)-g_l(a))ds$. Then, $G_l(x)$ exists and $G_l(x)\ge0$ for all  $x\in\mathbb{R}$, $G_{l}(x)=0$ if and only if $g_l(x)=g_l(a)$, and $G_l(x)\rightarrow\infty$ as $|x|\rightarrow\infty$.
\end{lemma}

\section{EVENT-TRIGGERED CONTROL}
\label{secmain}
In this section, we consider the multi-agent system (\ref{system}) defined over a digraph $\mathcal G$.
In the literature, the following distributed consensus protocol is often considered \cite{lim2016consensus,wang2016conditions},
\begin{align}\label{inputc}
u_i(t)=-\sum_{j=1}^{n}L_{ij}y_j(t).
\end{align}
\begin{remark}\label{remark1}
To implement (\ref{inputc}), continuous-time outputs from neighbours are needed. However, it is often unpractical to require continuous communication in physical applications. Moreover, if the nonlinear function $g$ is set as the saturation function and at some time $t_0$, $|x_i(t_0)|>h$, then due to the continuity of $x_i(t)$, there exists $t_1>t_0$ such that $|x_i(t)|\ge h,\forall t\in[t_0,t_1]$. Then $y_i(t)=\sat_h(x_i(t))$ is a constant during $[t_0,t_1]$. So it is also waste to continuously transmit $y_i(t)$ during $[t_0,t_1]$ since no new information is provided.
\end{remark}

Inspired by the idea of event-triggered control for multi-agent systems \cite{dimarogonas2012distributed}, we use instead of (\ref{inputc}) the following event-triggered  consensus protocol
\begin{align}\label{input}
u_i(t)=-\sum_{j=1}^{n}L_{ij}y_j(t^j_{k_j(t)})=-\sum_{j=1}^{n}L_{ij}g(x_j(t^j_{k_j(t)})),
\end{align}
where $k_{j}(t)=\argmax_{k}\{t^{j}_{k}\le t\}$, the increasing time agent-wise sequence $\{t_{k}^{j}\}_{k=1}^{+\infty}$, $j\in[n]$, named triggering time sequence of agent $v_j$ which will be determined later. Without loss of generality, we assume $t^j_1=0, j\in[n]$. Note that the control protocol (\ref{input}) only updates at the triggering times and is constant between consecutive triggering times. For simplicity, let $\hat{x}_{i}(t)=x_{i}(t_{k_{i}(t)}^{i})$ and
$e_{i}(t)=g(\hat{x}_{i}(t))-g(x_{i}(t))$.

In the following subsections, we will design triggering time sequence such that consensus is achieved for the multi-agent system (\ref{system}) with event-triggered consensus protocol (\ref{input}).
We first consider the case that the underlying digraph is strongly connected in Section~\ref{sec-sc}. We then consider the general case that the underlying digraph has a directed spanning tree in Section~\ref{sec-DST}.

\subsection{Strongly Connected Digraphs}\label{sec-sc}
We begin with the case that the underlying digraph is strongly connected.
Before presenting our main results, the following lemma is useful.
\begin{lemma}\label{lemma2}
Consider the multi-agent system (\ref{system}) and (\ref{input}).
Let $\xi^{\top}=[\xi_{1},\dots,\xi_{n}]$  be the positive vector defined in Lemma \ref{lemmaL}.
The weighted average of all agents' states $\bar{x}(t)=\sum_{i=1}^{n}\xi_ix_i(t)=[\bar{x}_1(t),\dots,\bar{x}_p(t)]^{\top}$ is a constant vector in $\mathbb{R}^p$, i.e., $\bar{x}(t)=\bar{x}(0)$.
\end{lemma}
{\bf Proof}:  The result holds since it follows from (\ref{system}) and (\ref{input}) that the time derivative of the average value is given by
\begin{align*}
\dot{\bar{x}}(t)=&\sum_{i=1}^{n}\xi_i\dot{x}_i(t)=-\sum_{i=1}^{n}\xi_i\sum_{j=1}^{n}L_{ij}g(\hat{x}_j)\\
=&-\sum_{j=1}^{n}g(\hat{x}_j)\sum_{i=1}^{n}\xi_iL_{ij}=0,
\end{align*}
where the last equality holds since $\xi^{\top}L=0$.

We are now ready to present our first main result.
\begin{theorem}\label{statictheorem}
Consider the multi-agent system (\ref{system}) with even-triggered control protocol (\ref{input}).
Suppose that the first four parts in Assumption~\ref{assnonlinearg} hold and the underlying graph $\mathcal G$ is directed and strongly connected. Suppose $x(0)\neq {\bf1}_n\otimes \bar{x}(0)$.
Given $\alpha_i>0,\beta_i>0$ and the first triggering time $t^i_1=0$, agent $v_i$ determines the triggering times $\{t^i_k\}_{k=2}^{\infty}$ by
\begin{align}\label{statictriggersingle}
t^i_{k+1}=\max_{r\ge t^i_k}\Big\{r:\|e_i(t)\|^2\le \alpha_ie^{-\beta_it},
\forall t\in[t^i_k,r]\Big\}.
\end{align}
Then (i) there is no Zeno behavior; (ii) the consensus can be achieved if
\begin{align*}
|\bar{x}_l(0)|\le h,l\in[p].
\end{align*}
Moreover, the consensus value is $\bar{x}(0)$.
\end{theorem}
{\bf Proof}: (i) Firstly, we prove that there is no Zeno behavior by contradiction.
Suppose there exists Zeno behavior, then there exists an agent $v_i$, such that $\lim_{k\rightarrow+\infty}t^i_k=T_0$ for some constant $T_0$.

Noting that $x_i(t),~i\in[n]$ are continuous, we know that there exists a constant $K_1$ such that $\|x_i(t)\|\le K_1,~\forall t\in[0,T_0]$. Thus, there exists a constant $K_2$ such that $\|g(x_i(t))\|\le K_2,~\forall t\in[0,T_0]$ due to the first part of Assumption 1. Hence, $$\|u_i(t)\|\le2L_{ii}K_2.$$
Let $\mathcal{S}_1=\{x\in\mathbb{R}^p:\|x\|\le K_1\}$, then from the fourth part in Assumption~\ref{assnonlinearg}, we know that there exists a constant $K(\mathcal{S}_1)>0$ such that $\|g(x_i(t))-g(\hat{x}_i(t))\|\le K(\mathcal{S}_1)\|x_i(t)-\hat{x}_i(t)\|,~\forall t\in[0,T_0]$. Thus, we can conclude that for $t\in[0,T_0]$ one sufficient condition to guarantee the inequality in condition (\ref{statictriggersingle}) is
\begin{align}\label{suffi1}
\|f_i(t)\|:=\|\hat{x}_i(t)-x_i(t)\|\le\sqrt{\frac{\alpha_i}{K(\mathcal{S}_1)}}e^{-\frac{\beta_it}{2}}.
\end{align}
Again noting that $$\bigg|\frac{d\|f_i(t)\|}{dt}\bigg|\le\|\dot{x}_i(t)\|=\|u_i(t)\|\le2L_{ii}K_2,$$ and $\|\hat{x}_i(t^i_{k})-x_i(t^i_{k})\|=0$ for any triggering time $t^i_k$, we can conclude that one sufficient condition to condition (\ref{suffi1}) is
\begin{align}\label{suffi2}
(t-t^i_k)2L_{ii}K_2\le\sqrt{\frac{\alpha_i}{K(\mathcal{S}_1)}}e^{-\frac{\beta_it}{2}}.
\end{align}

Let $\varepsilon_0=\frac{\sqrt{\alpha_i}}{4L_{ii}K_2\sqrt{K(\mathcal{S}_1)}}e^{-\frac{1}{2}\beta_iT_0}>0$. Then from the property of limits,  there exists a positive integer $N(\varepsilon_0)$ such that
\begin{align}\label{zeno}
t^i_k\in[T_0-\varepsilon_0,T_0],~\forall k\ge N(\varepsilon_0).
\end{align}

Now suppose that the $N(\varepsilon_0)$-th triggering time of agent~$i$, $t^i_{N(\varepsilon_0)}$, has been determined. Let $t^i_{N(\varepsilon_0)+1}$ and $\tilde{t}^i_{N(\varepsilon_0)+1}$ denote the next triggering time determined by (\ref{statictriggersingle}) and (\ref{suffi2}), respectively.
Then, $t^i_{N(\varepsilon_0)+1}\ge\tilde{t}^i_{N(\varepsilon_0)+1}$, and
\begin{align*}
&t^i_{N(\varepsilon_0)+1}-t^i_{N(\varepsilon_0)}
\ge\frac{\sqrt{\alpha_i}}{2L_{ii}K_2\sqrt{K(\mathcal{S}_1)}}e^{-\frac{1}{2}\beta_it^i_{N(\varepsilon_0)+1}}\nonumber\\
&\ge\frac{\sqrt{\alpha_i}}{2L_{ii}K_2\sqrt{K(\mathcal{S}_1)}}e^{-\frac{1}{2}\beta_iT_0}=2\varepsilon_0,
\end{align*}
which contradicts to (\ref{zeno}). Therefore, there is no Zeno behavior.

(ii) Consider a  Lyapunov candidate:
\begin{align*}
W(t)=V(t)+2\sum_{i=1}^{n}\frac{\xi_iL_{ii}\alpha_i}{\beta_i}e^{-\beta_it},
\end{align*}
where
\begin{align*}
V(t)=\sum_{i=1}^{n}\xi_i\sum_{l=1}^{p}\int_{\bar{x}_l(0)}^{c_l(x_{i}(t))}(g_l(s)-g_l(\bar{x}_l(0)))ds.
\end{align*}
From Lemmas \ref{lemmaL}--\ref{lemma3}, we know $V(t)\ge0$ and $V(t)=0$ if and only if $g(x_i(t))=g(\bar{x}(0)),i\in[n]$.

The derivative of $V(t)$ along the trajectories of system (\ref{system}) and (\ref{input}) satisfies
\begin{align}
\dot{V}&(t)\nonumber\\
=&\sum_{i=1}^{n}\xi_i\sum_{l=1}^p[g_l(c_l(x_{i}(t)))-g_l(\bar{x}_l(0))]c_l(\dot{x}_{i}(t))\nonumber\\
=&-\sum_{i=1}^{n}\xi_i\sum_{l=1}^p[g_l(c_l(x_{i}(t)))-g_l(\bar{x}_l(0))]\sum_{j=1}^{n}L_{ij}g_l(c_l(\hat{x}_{j}(t)))\nonumber\\
=&-\sum_{i=1}^{n}\xi_i[g(x_i(t))]^\top\sum_{j=1}^{n}L_{ij}[g(x_j(t))+e_j(t)]\nonumber\\
&+\sum_{i=1}^{n}\xi_iL_{ij}[g(\bar{x}(0))]^\top\sum_{j=1}^{n}g(\hat{x}_j(t))\nonumber\\
\overset{*}{=}&-\sum_{i=1}^{n}\xi_i[g(x_i(t))]^\top\sum_{j=1}^{n}L_{ij}g(x_j(t))\nonumber\\
&-\sum_{i=1}^{n}\xi_i[g(x_i(t))]^\top\sum_{j=1}^{n}L_{ij}e_j(t)\nonumber\\
\overset{**}{=}&\sum_{i=1}^{n}-\xi_iq_i(t)-\sum_{i=1}^{n}\xi_i[g(x_i(t))]^\top\sum_{j=1}^{n}L_{ij}e_j(t)\nonumber\\
\overset{*}{=}&\sum_{i=1}^{n}-\xi_iq_i(t)+\sum_{i=1}^{n}\sum_{j=1}^{n}\xi_iL_{ij}[e_j(t)]^\top[g(x_j(t))-g(x_i(t))]\nonumber\\
=&\sum_{i=1}^{n}-\xi_iq_i(t)\nonumber\\
&+\sum_{i=1}^{n}\sum_{j=1,j\neq i}^{n}\xi_iL_{ij}[e_j(t)]^\top[g(x_j(t))-g(x_i(t))]\nonumber\\
\le&\sum_{i=1}^{n}-\xi_iq_i(t)-\sum_{i=1}^{n}\sum_{j=1,j\neq i}^{n}\xi_iL_{ij}\|e_j(t)\|^2\nonumber\\
&-\sum_{i=1}^{n}\sum_{j=1,j\neq i}^{n}\xi_iL_{ij}\frac{1}{4}\|g(x_{j}(t))-g(x_{i}(t))\|^2\nonumber\\
=&\sum_{i=1}^{n}-\xi_iq_i(t)-\sum_{i=1}^{n}\sum_{j=1,j\neq i}^{n}\xi_iL_{ij}\|e_j(t)\|^2\nonumber\\
&-\sum_{i=1}^{n}\sum_{j=1}^{n}\xi_iL_{ij}\frac{1}{4}\|g(x_{j}(t))-g(x_{i}(t))\|^2\nonumber\\
=&\sum_{i=1}^{n}-\frac{1}{2}\xi_iq_i(t)-\sum_{i=1}^{n}\sum_{j=1,j\neq i}^{n}\xi_iL_{ij}\|e_j(t)\|^2\nonumber\\
\overset{*}{=}&\sum_{i=1}^{n}-\frac{1}{2}\xi_iq_i(t)+\sum_{i=1}^{n}\xi_iL_{ii}\|e_i(t)\|^2,\label{dV}
\end{align}
where
\begin{align*}
q_{i}(t)=-\frac{1}{2}\sum_{j=1}^{n}L_{ij}\|g(x_{j}(t))-g(x_{i}(t))\|^2\ge0,
\end{align*}
and the equalities denoted by $\overset{*}{=}$ hold since $\xi^\top L=0$, the equality denoted by $\overset{**}{=}$ holds since
\begin{align*}
&\sum_{i=1}^{n}-\xi_iq_i(t)\nonumber\\
&=\sum_{i=1}^{n}\frac{1}{2}\sum_{j=1}^{n}\xi_iL_{ij}\|g(x_{j}(t))-g(x_{i}(t))\|^2\nonumber\\
&=\sum_{i=1}^{n}\frac{1}{2}\sum_{j=1}^{n}\xi_iL_{ij}\Big[\|g(x_{j}(t))\|^2
+\|g(x_{i}(t))\|^2\Big]\nonumber\\
&~~~-\sum_{i=1}^{n}\sum_{j=1}^{n}\xi_iL_{ij}[g(x_{j}(t))]^\top g(x_{i}(t))\nonumber\\
&=-\sum_{i=1}^{n}\sum_{j=1}^{n}\xi_iL_{ij}[g(x_{j}(t))]^\top g(x_{i}(t)),
\end{align*}
and the inequality holds since $ab\le a^2+\frac{1}{4}b^2$.

Then, the derivative of $W(t)$ along the trajectories of system (\ref{system}) and (\ref{input}) is
\begin{align*}
&\dot{W}(t)\nonumber\\
&=\dot{V}(t)-2\sum_{i=1}^{n}\xi_iL_{ii}\alpha_ie^{-\beta_it}\nonumber\\
&\le\sum_{i=1}^{n}-\frac{1}{2}\xi_iq_i(t)+\sum_{i=1}^{n}\xi_iL_{ii}\|e_i(t)\|^2
-2\sum_{i=1}^{n}\xi_iL_{ii}\alpha_ie^{-\beta_it}\nonumber\\
&=\sum_{i=1}^{n}-\frac{1}{2}\xi_iq_i(t)-\sum_{i=1}^{n}\xi_iL_{ii}\alpha_ie^{-\beta_it}\le0.
\end{align*}
Then, by LaSalle Invariance Principle, there exists $a\in\mathbb{R}^p$, such that
\begin{align*}
\lim_{t\rightarrow+\infty}g(x_i(t))=a,i\in[n].
\end{align*}
This is equivalent to $\lim_{t\rightarrow+\infty}x_i(t)=b,i\in[n]$ with $g(b)=a$. Otherwise, from the second part in Assumption~\ref{assnonlinearg}, we know that either $x_i(t)\ge h{\bf1}_p$ for all $i\in[n]$ and $x_i(t)\neq h{\bf1}_p$ for some $i\in[n]$ or $x_i(t)\le h{\bf1}_p$ for all $i\in[n]$ and $x_i(t)\neq h{\bf1}_p$ for some $i\in[n]$. However, both cases contradict the condition $|\bar{x}_l(t)|=|\bar{x}_l(0)|\le h, l\in[p]$. Then, from $\bar{x}(t)=\sum_{i=1}^{n}\xi_ix_i(t)=\bar{x}(0)$, we have
\begin{align*}
\lim_{t\rightarrow+\infty}x_i(t)=\bar{x}(0),i\in[n].
\end{align*}

\begin{remark}
The event-triggered consensus protocol (\ref{input}) together with the triggering law (\ref{statictriggersingle}) is called event-triggered controller which is fully distributed. That is, each agent only requires its own state information and its neighbors' state information, without any a priori knowledge of global parameters, such as the eigenvalue of the Laplacian matrix. Moreover, one can easily check that by this event-triggered controller, there does not exist the waste mentioned in Remark \ref{remark1}.
\end{remark}

\begin{remark}
In the triggering law (\ref{statictriggersingle}), $\alpha_i$ and $\beta_i$ are design parameters than can be arbitrarily chosen by agent $v_i$. Intuitively, one can conclude that the larger $\alpha_i$ and the smaller $\beta_i$ the larger the inter-event time. This is also consistent with the definition of $\varepsilon_0$ in the proof. How do those design parameters $\alpha_i$ and $\beta_i$ affect the inter-event times and decay rate in  general is unclear. We leave this as future study.
\end{remark}

\subsection{Directed Spanning Trees}\label{sec-DST}
In this section, we extend the result in Theorem \ref{statictheorem} for the case where the underlying digraph is strongly connected to the general case where the underlying digraph has a directed spanning tree. 

The following mathematical methods are inspired by \cite{chen2007pinning,yi2019distributed}. By proper
permutation, we rewrite $L$ as the following Perron-Frobenius form:
\begin{eqnarray}
L=\left[\begin{array}{llll}L^{1,1}&L^{1,2}&\cdots&L^{1,M}\\
0&L^{2,2}&\cdots&L^{2,M}\\
\vdots&\vdots&\ddots&\vdots\\
0&0&\cdots&L^{M,M}
\end{array}\right]\label{PF}
\end{eqnarray}
where $L^{m,m}$ is with dimension $n_{m}$ and associated with the $m$-th
strongly connected component (SCC) of the digraph $\mathcal G$, denoted by $\SCC_{m}$,
$m\in[M]$.

If the digraph $\mathcal G$ has a directed spanning tree, then each $L^{m,m}$ is irreducible or
has one dimension and for each $m<M$, $L^{m,q}\ne 0$ for at least one
$q>m$. Define an auxiliary matrix
$\tilde{L}^{m,m}=[\tilde{L}^{m,m}_{ij}]_{i,j=1}^{n_{m}}$ as
\begin{eqnarray*}
\tilde{L}^{m,m}_{ij}=\begin{cases}L^{m,m}_{ij}&i\ne j\\
-\sum_{r=1,r\not=i}^{n_{m}}L^{m,m}_{ir}&i=j\end{cases}
\end{eqnarray*}
Then, let $D^{m}=L^{m,m}-\tilde{L}^{m,m}=\diag([D^{m}_{1},\dots,D^{m}_{n_{m}}])$, which is a diagonal semi-positive definite matrix and has at least one diagonal positive (nonzero).
Let ${\xi^{m}}^{\top}=[\xi^m_1,\dots,\xi^m_{n_m}]^\top$ be the positive left eigenvector of the irreducible
$\tilde{L}^{m,m}$ corresponding to the eigenvalue zero and has the sum of components equaling to $1$.
Denote $\Xi^{m}=\diag[\xi^{m}]$, $Q^{m}=\frac{1}{2}[\Xi^{m}L^{m,m}+(\Xi^{m}L^{m,m})^{\top}]$, $m=1,\dots,M$, and $U^{M}=\Xi^{M}-\xi^{M}(\xi^{M})^{\top}$.
Then, we have
\begin{lemma}\label{lemmaQ}
Under the setup above, $Q^{m}$ is positive definite for all $m<M$. $Q^M$ and $U^M$ are semi-positive definite.
Moreover,
\begin{align}\label{QU}
Q^M\ge \frac{\rho_2(Q^M)}{\rho(U^M)}U^M.
\end{align}
\end{lemma}
{\bf Proof}: (i) Consider a decomposition of the Euclidean space $\R^{n_m}$. Define
$
\mathcal S_{n_m}=\{x\in\R^{n_m}:~x_{i}=x_{j}~\forall~i,j\in[n_m]\}$ and
$
\mathcal L_{\xi^m}=\{x\in\R^{n_m}:~\sum_{i=1}^{n_m}\xi^m_{i}x_{i}=0\}
$
for the positive vector $\xi^m\in\R^{n_m}$. In this way, we can decompose $\R^{n_{m}}=\mathcal S_{n_{m}}\oplus\mathcal L_{\xi^{m}}$. For any $y\in\R^{n_{m}}$ and $x\ne 0$, we can find a unique decomposition of $y=y_{S}+y_{L}$ such that $y_{S}\in\mathcal S_{n_{m}}$ and $y_{L}\in\mathcal L_{\xi^{m}}$. Then,
\begin{eqnarray*}
y^{\top}\Xi^{k}L^{m,m}y=y^{\top}\Xi^{m}\tilde{L}^{m,m}y+y^{\top}\Xi^{m}D^{m}y.
\end{eqnarray*}
From Lemma \ref{lemmaL}, if $y_{L}\ne 0$, then $y^{\top}\Xi^{m}\tilde{L}^{m,m}y>0$; otherwise, $y=y_{S}=\alpha{\bf 1}_{n_m}$ for some $\alpha\ne 0$, then we have
\begin{eqnarray*}
y^{\top}\Xi^{m}D^{m}y=\sum_{i=1}^{n_{m}}D^{m}_{i}\xi^{m}_{i}\alpha^{2}>0.
\end{eqnarray*}
Therefore, we have $y^{\top}\Xi^{m}L^{m,m}y>0$ in any cases, which implies that $Q^m$ is positive definite. This completes the proof.

(ii) The proof of $Q^M$ and $U^M$ are semi-positive definite is straightforward. We thus omit the details here. The proof of (\ref{QU}) is same to explanation after Lemma 1 in \cite{yi2016distributed}.

Let $N_0=0,~N_{m}=\sum_{i=1}^{m}n_{i},m\in[M]$. Then the $i$-th agent in $\SCC_m$ is the $N_{m-1}+i$-th agent in the whole graph. In the following, we exchangeably use $v^m_i$ and $v_{N_{m-1}+i}$ to denote this agent. Accordingly, denote $x^m_i(t)=x_{N_{m-1}+i}(t)$, $\hat{x}^m_i(t)=\hat{x}_{N_{m-1}+i}(t)$ and define $x^{m}(t)=[{x^{m^{\top}}_1}(t),\dots,{x^{m^{\top}}_{n_m}}(t)]^{\top}$.
Let $\nu(t)=[\nu_1(t),\dots,\nu_p(t)]^\top=\sum_{r=1}^{n_{M}}\xi^{M}_{l}x^{M}_{l}(t)\in\mathbb{R}^p$.
Then from Lemma \ref{lemma2} we know that $\nu(t)\equiv\nu(0)$.

Our second main result is given in the following theorem.
\begin{theorem}\label{dynamictheorem}
Consider the multi-agent system (\ref{system}) and (\ref{input}).
Suppose that Assumption~\ref{assnonlinearg} holds and the  underlying digraph $\mathcal G$ has a spanning tree, and $L$ is written in the form of (\ref{PF}). Suppose $x(0)\neq {\bf1}_n\otimes \nu(0)$. Given $\alpha_i>0,\beta_i>0$ and the first triggering time $t^i_1=0$, agent $v_i$ determines the triggering times $\{t^i_k\}_{k=2}^{\infty}$ by the triggering law (\ref{statictriggersingle}).
Then (i) there is no Zeno behavior; (ii) the consensus can be achieved if
\begin{align*}
|\nu_l(0)|< h,l\in[p].
\end{align*}
Moreover, the consensus value is $\nu(0)$.
\end{theorem}
{\bf Proof}: (i) This part of the proof is the same as the proof of its counterpart in Theorem \ref{statictheorem}. Thus, we omit it here.



(ii) For simplicity, hereby we only consider the case of $M=2$. The case $M>2$ can be treated in the same way.

Firstly, let's consider $\SCC_2$. All agents in $\SCC_2$ do not dependent on any agents in $\SCC_1$. Thus, $\SCC_2$ has a strongly connected underlying digraph. Then from Theorem~\ref{statictheorem}, we have
\begin{align*}
\lim_{t\rightarrow+\infty}x_i^2(t)=\nu(0),i\in[n_2].
\end{align*}
Then, from $x^2_{i,l}(t),i\in[n_2],l\in[p]$ are continuous and $|\nu_l(0)|<h$, we can conclude that there exists a constant $T_1\ge0$ such that
\begin{align}\label{state}
|c_l(x^2_{i}(t))|<h,\forall t\ge T_1,i\in[n_2],l\in[p].
\end{align}

Secondly, let's consider $\SCC_1$.
Similar to $V(t)$, define
\begin{align*}
V_1(t)&=\sum_{i=1}^{n_1}\xi^1_i\sum_{l=1}^{p}\int_{\nu_l(0)}^{c_l(x^1_{i}(t))}[g_l(s)-g_l(\nu_l(0))]ds,\\
V_2(t)&=\sum_{i=1}^{n_2}\xi^2_i\sum_{l=1}^{p}\int_{\nu_l(0)}^{c_l(x^2_{i}(t))}[g_l(s)-g_l(\nu_l(0))]ds.
\end{align*}
From Lemmas \ref{lemmaL}--\ref{lemma3}, $V_1(t)\ge0$ and $V_2(t)\ge0$.

Similar to (\ref{dV}), the derivative of $V_2(t)$ along the trajectories of system (\ref{system}) and (\ref{input}) satisfies
\begin{align*}
\frac{dV_2(t)}{dt}\le&
\sum_{i=1}^{n_2}\sum_{j=1}^{n_2}\frac{1}{4}\xi^2_iL^{2,2}_{ij}
\|g(x^2_j(t))-g(x^2_i(t))\|^2\nonumber\\
&+\sum_{i=1}^{n_2}\xi^2_iL^{2,2}_{i,i}\|e^2_i(t)\|^2
\end{align*}
Then, for all $t\ge T_1$, we know that
\begin{align}\label{dv2}
&\frac{dV_2(t)}{dt}\nonumber\\
&\le
\sum_{i=1}^{n_2}\sum_{j=1}^{n_2}\frac{1}{4}\xi^2_iL^{2,2}_{ij}
\|g(x^2_j(t))-g(x^2_i(t))\|^2\nonumber\\
&~~~+\sum_{i=1}^{n_2}\xi^2_iL^{2,2}_{i,i}\|e^2_i(t)\|^2\nonumber\\
&\le
\sum_{i=1}^{n_2}\sum_{j=1}^{n_2}\frac{(\varrho)^2}{4}\xi^2_iL^{2,2}_{ij}
\|x^2_j(t)-x^2_i(t)\|^2+\sum_{i=1}^{n_2}\xi^2_iL^{2,2}_{i,i}\|e^2_i(t)\|^2\nonumber\\
&=-\frac{(\varrho)^2}{4}[x^2(t)]^\top(Q^2\otimes I_p)x^2(t)
+\sum_{i=1}^{n_2}\xi^2_iL^{2,2}_{i,i}\|e^2_i(t)\|^2\nonumber\\
&\le-\frac{(\varrho)^2\rho_2(Q^2)}{4\rho(U^2)}[x^2(t)]^\top(U^2\otimes I_p)x^2(t)\nonumber\\
&~~~+\sum_{i=1}^{n_2}\xi^2_iL^{2,2}_{i,i}\|e^2_i(t)\|^2\nonumber\\
&=-\frac{(\varrho)^2\rho_2(Q^2)}{2\rho(U^2)}\mu(t)+\sum_{i=1}^{n_2}\xi^2_iL^{2,2}_{i,i}\|e^2_i(t)\|^2,
\end{align}
where the second inequality holds since (\ref{state}) as well as the last part in Assumption~\ref{assnonlinearg} and the last inequality holds since Lemma~\ref{lemmaQ} and
\begin{align*}
\mu(t)=\frac{1}{2}\sum_{i=1}^{n_2}\xi_i^2\|g(x^2_i(t))-\nu(0)\|^2\ge0.
\end{align*}

The derivative of $V_1(t)$ along the trajectories of system (\ref{system}) and (\ref{input}) satisfies
\begin{align}\label{dv1}
&\frac{dV_1(t)}{dt}=\sum_{i=1}^{n_1}\xi_i^1\sum_{l=1}^{p}[g_l(c_l(x^1_{i}(t)))-g_l(\nu_l(0))]c_l(\dot{x}^1_{i}(t))\nonumber\\
&=\sum_{i=1}^{n_1}\xi_i^1[g(x^1_i(t))-g(\nu(0))]^\top\dot{x}^1_i(t)\nonumber\\
&=\sum_{i=1}^{n_1}\xi_i^1[g(x^1_i(t))-g(\nu(0))]^\top\Big\{
-\sum_{j=1}^{n_1}L^{1,1}_{ij}g(\hat{x}_j^1(t))\nonumber\\
&~~~-\sum_{j=1}^{n_2}L^{1,2}_{ij}g(\hat{x}_j^2(t))\Big\}\nonumber\\
&=\sum_{i=1}^{n_1}\xi_i^1[g(x^1_i(t))-g(\nu(0))]^\top\nonumber\\
&~~~\times\Big\{-\sum_{j=1}^{n_1}L^{1,1}_{ij}[g(x_j^1(t))+e^1_j(t)]\nonumber\\
&~~~-\sum_{j=1}^{n_2}L^{1,2}_{ij}[g(x_j^2(t))+e^2_j(t)]\Big\}\nonumber\\
&=\sum_{i=1}^{n_1}\xi_i^1[g(x^1_i(t))-g(\nu(0))]^\top\nonumber\\
&~~~\times\Big\{-\sum_{j=1}^{n_1}L^{1,1}_{ij}[g(x_j^1(t))-g(\nu(0))]-\sum_{j=1}^{n_1}L^{1,1}_{ij}e^1_j(t)\nonumber\\
&~~~-\sum_{j=1}^{n_2}L^{1,2}_{ij}[g(x_j^2(t))-g(\nu(0))]-\sum_{j=1}^{n_2}L^{1,2}_{ij}e^2_j(t)\Big\}\nonumber\\
&=-[g(x^1(t))-{\bf1}_{n_1}\otimes g(\nu(0))]^\top[Q^1\otimes I_{n_1}]\nonumber\\
&~~~\times[g(x^1(t))-{\bf1}_{n_1}\otimes g(\nu(0))]\nonumber\\
&~~~-\sum_{i=1}^{n_1}\xi_i^1[g(x^1_i(t))-g(\nu(0))]^\top\nonumber\\
&~~~\times\Big\{\sum_{j=1}^{n_2}L^{1,2}_{ij}[g(x_j^2(t))-g(\nu(0))]\nonumber\\
&~~~+\sum_{j=1}^{n_1}L^{1,1}_{ij}e^1_j(t)+\sum_{j=1}^{n_2}L^{1,2}_{ij}e^2_j(t)\Big\}\nonumber\\
&\le-\rho(Q^1)\sum_{i=1}^{n_1}\|g(x^1_i(t))-g(\nu(0))\|^2\nonumber\\
&~~~+\frac{\rho(Q^1)}{2}\sum_{i=1}^{n_1}\|g(x^1_i(t))-g(\nu(0)))\|^2\nonumber\\
&~~~+\frac{1}{\rho(Q^1)}\sum_{i=1}^{n_1}\Big\|\xi_1^1\sum_{j=1}^{n_2}L^{1,2}_{ij}[g(x_j^2(t))-g(\nu(0))]\Big\|^2\nonumber\\
&~~~+\frac{1}{\rho(Q^1)}\sum_{i=1}^{n_1}\Big\|\xi_1^1\sum_{j=1}^{n_1}L^{1,1}_{ij}e^1_j(t)
+\xi_1^1\sum_{j=1}^{n_2}L^{1,2}_{ij}e^2_j(t)\Big\|^2\nonumber\\
&\le-\frac{\rho(Q^1)}{2}\sum_{i=1}^{n_1}\|g(x^1_i(t))-g(\nu(0))\|^2\nonumber\\
&~~~+n_2\sum_{j=1}^{n_2}d^2_j\|g(x_j^2(t))-g(\nu(0))\|^2\nonumber\\
&~~~+2n_1\sum_{j=1}^{n_1}d^1_j\|e^1_j(t)\|^2+2n_2\sum_{j=1}^{n_2}d^2_j\|e^2_j(t)\|^2.
\end{align}
where $$d^1_j=\frac{\sum_{i=1}^{n_1}[\xi^1_iL^{1,1}_{ij}]^2}{\rho(Q^1)},d^2_j=\frac{\sum_{i=1}^{n_1}[\xi^1_iL^{1,2}_{ij}]^2}{\rho(Q^1)}.$$

Let $\mathcal{S}_2=[-h,h]$, then from the fourth part in Assumption~\ref{assnonlinearg}, we know that there exists $K(\mathcal{S}_2)>0$ such that $|g_l(x)-g_l(y)|\le K(\mathcal{S}_2)|x-y|$ for all $x,y\in\mathcal{S}_2$. Noting (\ref{state}), we have
\begin{align}
&n_2\sum_{j=1}^{n_2}d^2_j\|g(x_j^2(t))-g(\nu(0))\|^2\nonumber\\
&\le n_2(K(\mathcal{S}_2))^2\sum_{j=1}^{n_2}d^2_j\|x_j^2(t)-\nu(0)\|^2\nonumber\\
&\le\frac{2n_2(K(\mathcal{S}_2))^2\max\{d^2_j\}}{\min\{\xi^2_j\}}\mu(t),\forall t\ge T_1.
\end{align}

The triggering law  (\ref{statictriggersingle}) yields
\begin{align}\label{e12}
\|e^1_j(t)\|^2\le\alpha^1_je^{-\beta^1_jt},\|e^2_j(t)\|^2\le\alpha^2_je^{-\beta^2_jt},\forall t\ge0.
\end{align}

Consider a Lyapunov candidate:
\begin{align*}
&W_r(t)=V_1(t)+K_vV_2(t)+3n_1\sum_{j=1}^{n_1}\frac{d^1_j\alpha^1_j}{\beta^1_j}e^{-\beta^1_jt}\nonumber\\
&+3n_2\sum_{j=1}^{n_2}\frac{d^2_j\alpha^2_j}{\beta^2_j}e^{-\beta^2_jt}
+K_v\sum_{j=1}^{n_2}\frac{\xi^2_jL^{2,2}_{jj}\alpha^2_j}{\beta^2_j}e^{-\beta^2_jt},
\end{align*}
where $$K_v=\Big(\frac{2n_2(K(\mathcal{S}_2))^2\max\{d^2_j\}}{\min\{\xi^2_j\}}+1\Big)\frac{2\rho(U^2)}{(\varrho)^2\rho_2(Q^2)}.$$
The derivative of $W_r(t)$ along the trajectories of system (\ref{system}) and (\ref{input}) is
\begin{align}\label{dwr}
&\frac{dW_r(t)}{dt}=\dot{V}_1(t)+K_v\dot{V}_2(t)-3n_1\sum_{j=1}^{n_1}d^1_j\alpha^1_je^{-\beta^1_jt}\nonumber\\
&-3n_2\sum_{j=1}^{n_2}d^2_j\alpha^2_je^{-\beta^2_jt}
-K_v\sum_{j=1}^{n_2}\xi^2_jL^{2,2}_{jj}\alpha^2_je^{-\beta^2_jt}.
\end{align}
Then, from (\ref{dv2})--(\ref{dwr}), for any $t\ge T_1$, we have
\begin{align}\label{dwr2}
\frac{dW_r(t)}{dt}
\le&-\frac{\rho(Q^1)}{2}\sum_{i=1}^{n_1}\|g(x^1_i(t))-g(\nu(0))\|^2-\mu(t)\nonumber\\
&-n_1\sum_{j=1}^{n_1}d^1_j\alpha^1_je^{-\beta^1_jt}-n_2\sum_{j=1}^{n_2}d^2_j\alpha^2_je^{-\beta^2_jt}.
\end{align}
Then, by LaSalle Invariance Principle, similar to the proof in Theorem \ref{statictheorem}, we have
\begin{align*}
\lim_{t\rightarrow+\infty}x_j^1(t)=\lim_{t\rightarrow+\infty}x_i^2(t)=\nu(0),
\end{align*}
for all $j\in[n_1],i\in[n_2]$.

\begin{remark}
Different from Theorem~\ref{statictheorem} where $|\nu_l(0)|\le h,l\in[p]$ is sufficient to that the multi-agent system (\ref{system}) and (\ref{input}) under the  triggering law  (\ref{statictriggersingle}) achieves consensus, for Theorem~\ref{dynamictheorem}, $|\nu_l(0)|< h,l\in[p]$ is sufficient.
\end{remark}

\begin{remark}\label{remark-nec}
If $g_l(x)=g_l(h)$ when $x\ge h$ and $g_l(x)=g_l(-h)$ when $x\le -h$, then $|\bar{x}_l(0)|\le h, \; l\in[p]$ is the necessary condition for achieving consensus. This assumption is satisfied by the saturation function.
The necessity can be proved by the proof of Theorem 3.1 in \cite{lim2016consensus} with some minor modifications. We thus omit the proof here.
\end{remark}

\section{SIMULATIONS}\label{secsimulation}

\begin{figure}[hbt]
\centering
\includegraphics[width=0.45\textwidth]{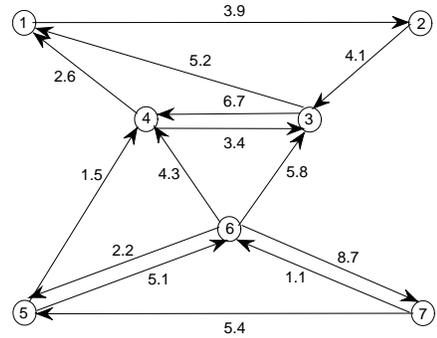}
\caption{The underlying digraph.}
\label{fig:1}
\end{figure}

\begin{figure}
\begin{subfigure}{.5\textwidth}
  \centering
  \includegraphics[width=.9\linewidth]{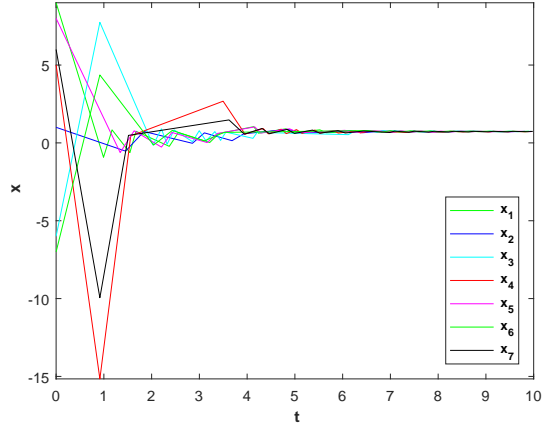}
  \caption{}
  \label{fig:2a}
\end{subfigure}%
\\
\begin{subfigure}{.5\textwidth}
  \centering
  \includegraphics[width=.9\linewidth]{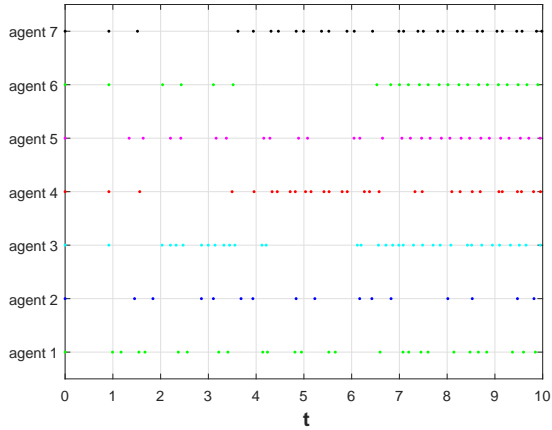}
  \caption{}
  \label{fig:2b}
\end{subfigure}
\caption{(a) The state evolutions of the multi-agent system (\ref{system}) and (\ref{input}) under the triggering law (\ref{statictriggersingle}).  (b) The triggering times for each agent.}
\label{fig:2}
\end{figure}

In this section, a numerical example is given to demonstrate the theoretical results.
Consider the case when the nonlinear function $g_l=\sat_h$ with $h=1$.
Consider a directed graph of seven agents with the Laplacian matrix
\begin{eqnarray*}
L=\left[\begin{array}{rrrrrrr}7.8& 0  &  -5.2 &  -2.6 &  0   & 0  &  0\\
    -3.9  &3.9 &0    & 0    & 0   & 0   & 0\\
    0     &-4.1 &13.3 &-3.4  & 0   &-5.8 &  0\\
    0    & 0   & -6.7  &12.5 &-1.5 & -4.3&  0\\
    0    & 0   & 0     &0    &7.6 &-2.2  &-5.4\\
    0    & 0   & 0     &0    &-5.1  &6.2 &-1.1\\
    0    & 0   & 0    & 0    &0    &-8.7  &8.7
\end{array}\right],
\end{eqnarray*}
whose topology is shown in Fig. \ref{fig:1}. The seven agents can be divided into two strongly connected components, i.e. the first four agents form a strongly connected component and the rest form anther.
The initial value of each agent is selected within the interval $[-10,10]$. First, we use $x(0)=[9, 1,   -6,    5,    8,   -7,   6]^{\top}$, then the weighted average initial states is $\nu(0)=0.7345$. Thus the sufficient condition
$|\nu(0)|< h$ in Theorem~\ref{dynamictheorem} is satisfied.
Fig. \ref{fig:2} (a) shows the state evolutions of the multi-agent system (\ref{system}) and (\ref{input}) under the  triggering law (\ref{statictriggersingle}) with $\alpha_i=\beta_i=10$. Fig. \ref{fig:2} (b) shows the corresponding triggering times for each agent. It can
be seen that  consensus is achieved.


\begin{figure}[hbt]
\centering
\includegraphics[width=0.45\textwidth]{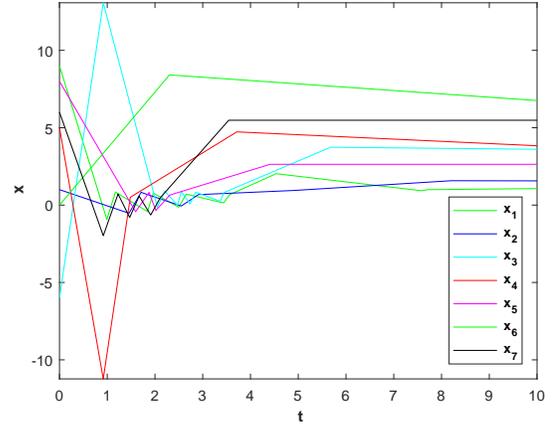}
\caption{The state evolutions of the multi-agent system (\ref{system}) and (\ref{input}) under the triggering law (\ref{statictriggersingle}).}
\label{fig:6}
\end{figure}

Then, we use $x(0)=[9,    1,   -6,   5,    8,   0,   6]^{\top}$, then the weighted average initial states is $\nu(0)=3.8962$.
Thus the necessary condition $|\nu(0)| \leq h$ for achieving consensus as stated Remark~\ref{remark-nec} is not satisfied.
Fig. \ref{fig:6} shows the state evolutions of the multi-agent system (\ref{system}) and (\ref{input}) under the  triggering law  (\ref{statictriggersingle}) with $\alpha_i=\beta_i=10$. It can
be seen that consensus is not achieved in this case since the weighted average initial state is not within the saturation level.

\section{CONCLUSION}\label{secconclusion}
In this paper, we developed a distributed event-triggered control law for multi-agent systems with nonlinear output under directed communication topologies.
We proposed different sufficient conditions related to the nonlinear function and the initial states to guarantee consensus is achieved when the underlying digraph is strongly connected or
has a directed spanning tree.
In addition, our event-triggered controller was shown to be free of Zeno behavior.
Future research directions include considering more general systems such as double integrator systems, time delays, and self-triggered control.

\bibliographystyle{IEEEtran}
\bibliography{refs}

\end{document}